\documentclass[
 reprint,
 amsmath,amssymb,
 aps,
 prl
]{revtex4-1}

\usepackage{graphicx}% Include figure files
\usepackage{dcolumn}% Align table columns on decimal point
\usepackage{bm}% bold math

\newcommand{\ket}[1]{|#1\rangle}

\begin{document}

\title{Scalable Spatial Super-Resolution using Entangled Photons}

\author{Lee A. Rozema$^{1\ast\dagger}$}

\author{James D. Bateman$^{1\dagger}$}
 
\author{Dylan H. Mahler$^1$}

\author{Ryo Okamoto$^{1,2,3}$}

\author{Amir Feizpour$^1$}

\author{Alex Hayat$^{1,4,5}$}

\author{Aephraim M. Steinberg$^{1,5}$}

\affiliation{
$^1$Centre for Quantum Information \& Quantum Control and Institute for Optical Sciences,
Dept. of Physics, 60 St. George St., University of Toronto, Toronto, Ontario, Canada M5S 1A7\\
$^2$Research Institute for Electronic Science, Hokkaido University, Kita-ku, Sapporo, Japan\\
$^3$The Institute of Scientific and Industrial Research, Osaka University, Mihogaoka 8-1, Ibaraki, Osaka, Japan\\
$^4$Department of Electrical Engineering, Technion, Haifa 32000, Israel\\
$^5$Canadian Institute for Advanced Research, Toronto, Ontario M5G 1Z8, Canada\\
$^\ast$Correspondence to:  lrozema@physics.utoronto.ca.\\
$^\dagger$These authors contributed equally to this work.
}

\date{\today}
\begin{abstract}
N00N states -- maximally path-entangled states of N photons -- exhibit spatial interference patterns sharper than any classical interference pattern. This is known as super-resolution. However, even with perfectly efficient number-resolving detectors, the detection efficiency of all previously demonstrated methods to measure such interference decreases exponentially with the number of photons in the N00N state, often leading to the conclusion that N00N states are unsuitable for spatial measurements.  Here, we create spatial super-resolution fringes with two-, three-, and four-photon N00N states, and demonstrate a scalable implementation of the so-called ``optical centroid measurement'' which provides an in-principle perfect detection efficiency.  Moreover, we compare the N00N-state interference to the corresponding classical super-resolution interference.  Although both provide the same increase in spatial frequency, the visibility of the classical fringes decreases exponentially with the number of detected photons, while the visibility of our experimentally measured N00N-state super-resolution fringes remains approximately constant with N.  Our implementation of the optical centroid measurement is a scalable method to measure high photon-number quantum interference, an essential step forward for quantum-enhanced measurements, overcoming what was believed to be a fundamental challenge to quantum metrology.
\end{abstract}
\maketitle

Many essential techniques in modern science and technology, from precise position sensing to high-resolution imaging to nanolithography, rely on the creation and detection of the finest possible spatial interference fringes using light.  Classically, all such measurements face a fundamental barrier related to the ``diffraction limit,'' which is determined by the wavelength of the light \cite{lord_rayleigh_investigations_1879}, but quantum entanglement can be used to surpass this limit by making the spatial interference fringes sharper (a result referred to as super-resolution) \cite{boto_quantum_2000, kok_quantum-interferometric_2001}.  In particular, the N-photon entangled ``N00N" state can display an interference pattern N times finer than that of classical light \cite{mitchell_super-resolving_2004, walther_broglie_2004}.  However, N00N states suffer from a weakness that has made their advantage controversial: the probability of all N photons arriving at the same place, and thus the detection efficiency, decreases exponentially with N \cite{steuernagel_concentration_2004, tsang_relationship_2007}.  Here we implement the optical centroid measurement (OCM) proposed by Tsang \cite{tsang_quantum_2009} to completely overcome this problem. A proof-of-principle experiment confirming the underlying concept of the OCM was recently performed \cite{shin_quantum_2011}, but, being limited to only two photons and two `movable' detectors, it could not probe the scaling properties nor demonstrate the efficiency gain of the OCM.  In our experiment, using an array of 11 fixed detectors, we measure two-, three-, and four-photon spatial fringes, and find that their visibility does not degrade with the number of entangled photons, clearly displaying the enhanced efficiency and scalability of the OCM.  The visibility of an unentangled OCM, on the other hand, decays exponentially.  In doing this, we have also achieved the highest spatial super-resolution to date \cite{dangelo_two-photon_2001, kawabe_quantum_2007, kim_observation_2011}. 

In 2000, Boto \textit{et al}. pointed out that entangled states of light offer a way to improve the resolution of interferometers beyond the diffraction limit, which determines the smallest spatial features achievable in classical optical systems \cite{boto_quantum_2000}.  This limit is set by the fringe spacing of the interference pattern created by two beams of wavelength $\lambda$ meeting at an angle $\theta$, which is $\lambda/(2\sin\frac{\theta}{2})$. Under no conditions can a spacing smaller than $\lambda/2$ be attained classically.  Boto \textit{et al.} circumvented this classical limit by introducing the entangled N00N state, an equal superposition of all N photons in mode $\vec{k_1}$ and all N photons in mode $\vec{k_2}$:
\begin{equation}
\label{eq:N00N}
\ket{\psi_N}=\frac{1}{\sqrt{2}}(\ket{N,0}_{\vec{k_1},\vec{k_2}}+\ket{0,N}_{\vec{k_1},\vec{k_2}}).
\end{equation}
If $\vec{k_1}$ and $\vec{k_2}$ are two spatial modes which interfere at a detection plane (figure 1a), the probability of detecting \textit{all} of the photons at a given position will display spatial fringes with a period of $\lambda/(2N\sin\frac{\theta}{2})$, where $\theta$ is the angle between $\vec{k_1}$ and $\vec{k_2}$ (assuming $|\vec{k_1}|=|\vec{k_2}|=\frac{2\pi}{\lambda}$).  The period of the N00N fringes is N times smaller than that of classical fringes, suggesting that N00N states could be used to increase the resolution of optical systems by a factor of N.  This observation has led to much subsequent work on N00N states \cite{mitchell_super-resolving_2004, walther_broglie_2004,nagata_beating_2007, shalm_squeezing_2009, hofmann_high-photon-number_2007, afek_high-noon_2010} and their application in tasks such as quantum lithography and quantum imaging \cite{dangelo_two-photon_2001, kawabe_quantum_2007, kim_observation_2011, nasr_demonstration_2003}.  However, the individual photons in the N00N state cannot be localized to better than $\lambda/2$, regardless of the narrow spatial scale of the N-photon correlation fringes \cite{steuernagel_concentration_2004, tsang_relationship_2007}.  This means that the probability of a given photon landing within some small region of size $r$ will always be $\lesssim 2r/\lambda$, and that the probability for all N photons to arrive at the same region is $\lesssim (2r/\lambda)^N$.  Thus the efficiency of such a detection scheme decreases exponentially with $N$, leading to the conclusion that N00N states are of little practical use for applications requiring spatial interference.  The OCM was proposed to address this problem, resurrecting the hope of applying such states to high-resolution position measurements \cite{tsang_quantum_2009}.

\begin{figure}
\includegraphics[scale=.42]{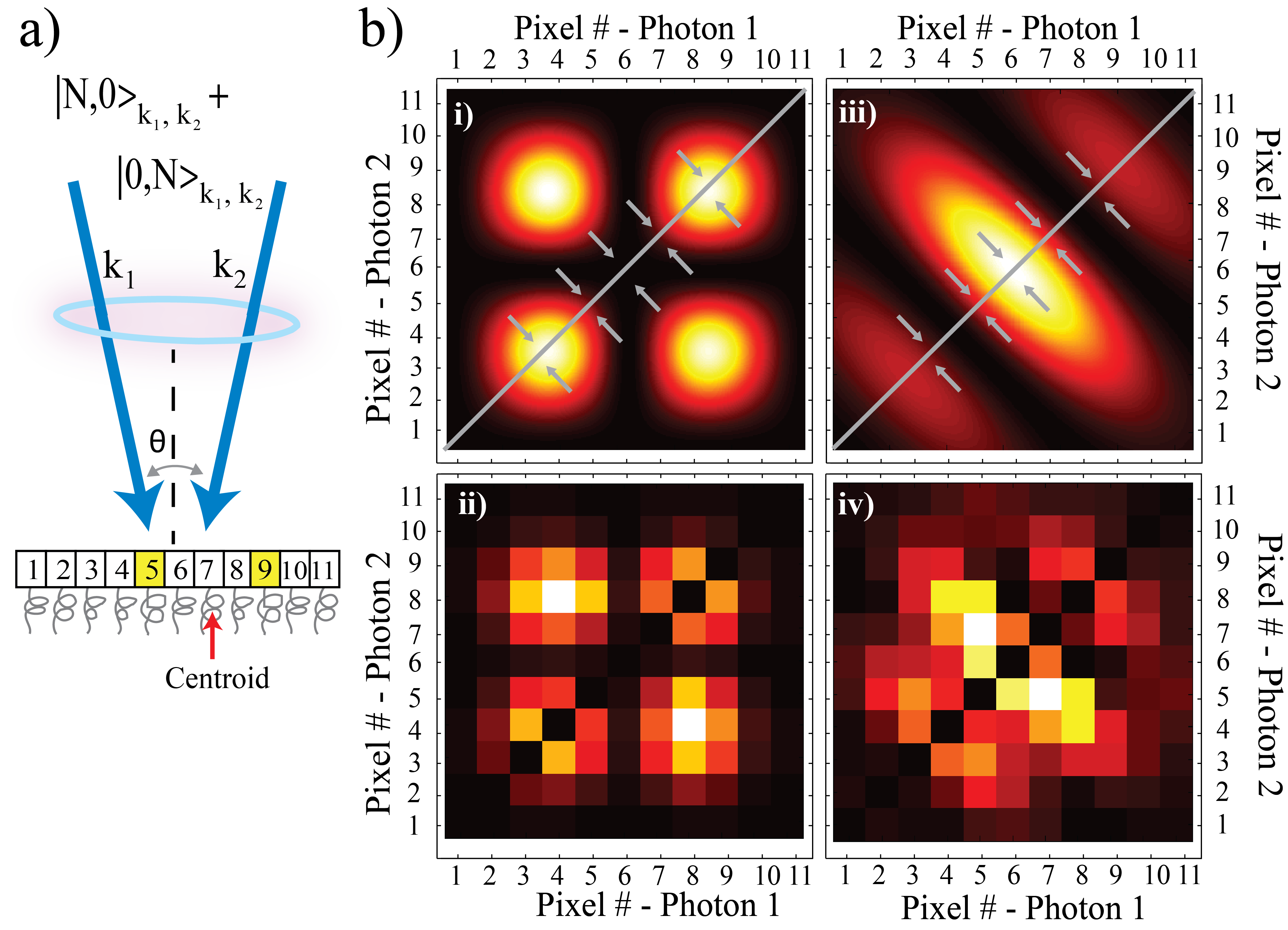}
\caption{\textbf{ The Centroid measurement:} \textit{a) Schematic of the detection scheme --} Light from two modes, $\vec{k_1}$ and $\vec{k_2}$, is incident on an array of single photon detectors. Correlations between all the detectors are measured and recorded. \textit{b) Two-photon joint-probability functions --} i) and iii) are the theoretical two-photon joint-probability functions when classical light and N=2 N00N states illuminate the detector array, respectively. A two-photon absorption measurement requires $x_1=x_2$, and as such it only samples the grey diagonal line, drawn in i) and iii), to produce a spatial interference pattern. This results in discarding most of the photon pairs.  The centroid samples the entire two-photon correlation function by \textit{projecting} all of the data onto the grey diagonal line, effectively detecting all of photon pairs. ii) and iv) are the experimental two-photon joint-probability functions.  Since there is no way to distinguish photon 1 from photon 2, we could only measure the half of the plot where $x_1>x_2$; in plots ii) and iv) we have mirrored this data about the diagonal ($x_1=x_2$) for comparison to theory.  The diagonal region is dark since two photons arriving simultaneously at the same SPCM cannot both be detected.}
\end{figure}

The OCM displays N-fold super-resolution without requiring all the photons to arrive at the same point in space.  Instead, it keeps track of every single N-photon event, regardless of which combination of detectors fires.  By using appropriate post-processing, the OCM nevertheless unveils the N-photon quantum interference.  For simplicity, consider interfering a two-photon N00N state on an array of photon detectors (figure 1a).  Most of the time, two different detectors fire; occasionally, both photons reach the same detector. The original proposals only retained those rare events in which a single detector registered both photons, observing that this rate exhibited sub-wavelength fringes as a function of detector position. By contrast, the OCM keeps \textit{all} the events, recording the ``centroid,'' or the average of the detected positions of the photons -- remarkably, sub-wavelength fringes are also observed as a function of this centroid, obviating the need to discard the bulk of the events \cite{tsang_quantum_2009}.

A more visual way to look at the OCM is illustrated in figure 1b, which shows plots of the joint probabilities for photon 1 to arrive at pixel $x_1$ and photon 2 to arrive at pixel $x_2$ in coincidence.  In a two-photon--absorber measurement, an event is only registered if $x_1=x_2$ (both photons arrive at the same point).  The resulting signal will be given by the photon correlations along the grey diagonal lines drawn in $i$ (for classical light) and $iii$) (for a N00N state).  This also means that all of the other possible events are discarded.  The OCM signal, on the other hand, utilizes all of this data.  For two photons, the OCM signal can be visualized as the integral of the joint-probability functions onto the grey diagonal lines (shown $i$ and $iii$). In other words, it is a histogram of all of the points plotted versus their centroid position, $(x_1+x_2)/2$.  Notice that the OCM also increases the number of ``effective pixels'' by a factor of N, since if, for example, detectors 3 and 4 fire the value of the centroid is $3.5$, making it possible to detect super-resolution features which are much smaller than the physical pixel size.  Similar joint-probability functions can be made for any value of N. In this case, the N-photon centroid is computed from an N-dimensional integral.  The N-photon OCM signal at position $X$ is number (rate) of events whose centroid, $X_c=(x_1+\dots+x_N)/N$, is equal to $X$, where $x_i$ is the pixel at which photon $i$ is detected \cite{gulfam_numerical_2013}.  Unlike the N-photon-absorbing proposal, which keeps only $O(\frac{1}{D^N})$ of the N-photon events (if there are D pixels), this method uses them all.

From figure 1b, it can be seen that performing an OCM on a N00N state will result in a signal with the same periodicity as the N-photon absorption detection scheme. It is also evident that performing the OCM on classical light will result in fringes with this enhanced periodicity.  Although the classical OCM indeed results in a signal with a period of $\lambda/(2N\sin\theta)$, its visibility decreases exponentially with $N$, as $V_c=\frac{1}{2^{N-1}}$.  (As shown in the Supplemental Material, this is because performing an N-fold OCM on classical light results in signal which is the convolution of the singles intensity pattern with itself N times.)  However, as we show experimentally for two-, three-, and four-photon N00N states, the visibility of the N00N state OCM remains constant, independent of $N$.

\begin{figure}
\includegraphics[scale=.48]{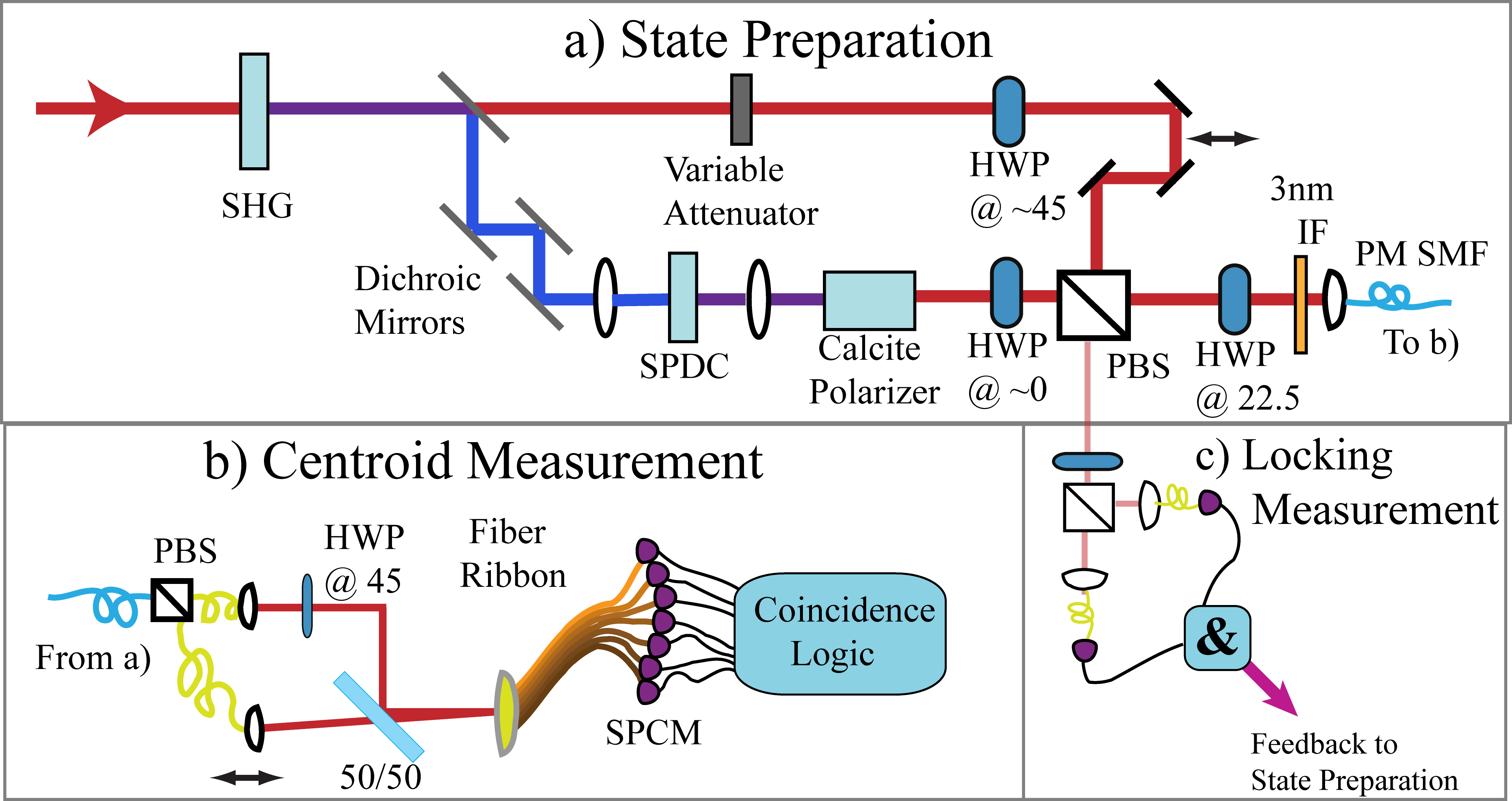}
\caption{{\bf Schematic of the experimental apparatus:} \textit{a) N00N state preparation -- } Laser light and light from type-I collinear down conversion are combined to create polarization N00N states for a range of different values of N. \textit{b) Centroid Measurement -- } The polarization N00N states are converted into path-entangled N00N states, and interfered on a multi-mode fiber ribbon connected to 11 single-photon counting modules (SPCM). \textit{c) Locking Measurement -- } A small amount of down-converted and laser light is sent into the ``locking port'', and used to measure any phase drift between the laser and down-conversion paths so that it can be corrected.}
\end{figure}

Experimentally, we produce N00N states in polarization using a relatively new technique which, given bright down-conversion sources, can produce N00N states of arbitrary $N$ \cite{hofmann_high-photon-number_2007, afek_high-noon_2010}.  Our source (figure 2a) creates N00N states by combining laser light with light from a Type-I collinear down-conversion (DC) source.  The DC light is generated by pumping a 1mm BBO crystal, cut for type-I phase matching, with 600mW average power of 404nm light. The 404nm pump is generated by frequency-doubling light from a femtosecond Ti:Sapph laser (running at 1.4W average power and 76 MHz repetition rate, and centered at 808nm) using a 2mm long crystal of BBO.  We measure approximately 12,000 two-photon counts/s from the DC source, with a coupling efficiency (pairs/singles) of $9.5\%$  when single-photon detectors are placed directly after the in-fiber polarizing beamsplitter (PBS) of panel b (including a diversion of $\approx3\%$ of the photon pairs to our locking measurement, to be discussed shortly).  The DC light is spatially overlapped with light from the Ti:Sapph laser at a PBS, passed through a half waveplate at $22.5^\circ$, and coupled into polarization-maintaining single-mode fiber.  To generate N00N states, the relative phase between the two arms must be set to zero, and the relative amplitudes balanced.  The source is optimized for N00N states of different N by simply changing the relative amplitude between the laser and DC light \cite{hofmann_high-photon-number_2007}.  Ideally, two- and three-photon N00N states can be made perfectly by setting the two-photon rate from the laser equal to that from the DC light (which we will refer to as configuration 1). Although our source cannot make perfect four-photon N00N states, it can, in principle, make a state which has $93\%$ fidelity with the ideal four-photon N00N state (the visibility of an OCM using this state should also be $93\%$). Making this four-photon state requires that the laser two-photon rate be three times larger than the DC two-photon rate.  However, even if the laser rate is further increased, increasing the four-photon count rate, the fidelity of the four-photon state will not be significantly degraded \cite{rosen_sub-rayleigh_2012}.  To make four-photon N00N states, we choose to make the laser two-photon rate 8.5 times larger than the DC rate (configuration 2), which leads to a fidelity with the ideal state of $85\%$, a theoretical OCM visibility of $85\%$, and increases the four-photon rate by about a factor of 10 (compared to the configuration in which the four-photon fidelity is optimized).

The phase between the two arms must also be stabilized in order to make N00N states.  We accomplished this by using a piezoelectric-driven trombone arm in the laser path.  To generate a feedback signal, small amounts of DC and laser light are sent through the other port of the state-preparation PBS, to the ``locking measurement'' (figure 2c).  We measure 500 down-converted two-photon counts/s ($\approx3\%$ of the detected down-converted light) at the locking measurement and 5000 two-photon counts/s from the laser, creating a low-fidelity two-photon N00N state. This state will be phase shifted if the phase between the two arms drifts, and can therefore be used to track the phase drift.  After acquiring counts for 5s, the phase of the state is measured, and then any phase drift between the two arms is corrected.  Using this locking mechanism, we are able to keep the N00N state source stable for days.

Once polarization N00N states are collected into polarization-maintaining single-mode fiber from the source, they are sent to the OCM apparatus (figure 2b). There, they are converted into path-entangled N00N states using an in-fiber PBS.  Once the output polarizations are matched, the two modes are spatially interfered, by overlapping them at a $50:50$ beamsplitter, at an angle of $\theta=0.16$ mrad.  These overlapped modes are focused onto a ``fiber-ribbon'' (which serves as our fixed-detector array) using cylindrical lenses. The visibility of the spatial interference pattern formed across the detector array is measured to be $90\%$ using classical light.  The phase of this spatial interference signal is actively locked by observing the phase on the detector array, and feeding this back to a piezoelectric actuator which moves one of the fiber collimators.

Our array of single-photon detectors consists of 12 multi-mode fibers, mounted in a standard MTP12 fiber ribbon.  The fiber cores are 62.5$\mu$m in diameter, and are mounted linearly with a 250$\mu$m separation between fiber centers.  This results in a fill factor (and therefore maximum coupling efficiency) of $\approx25\%$.  Eleven of these fibers are coupled to Perkin-Elmer single-photon counting modules (SPCM), which are all connected to a custom-built FPGA-based coincidence counting circuit capable of detecting coincidence events among all possible detector combinations.   Using classical light, we measure an overall coupling efficiency of $20\%$ into the 11 fibers making up our detector array.  The additional losses in the OCM apparatus further reduces the final coupling efficiency of the DC photons (pairs/singles) from $9.5\%$ to $0.2\%$ (measured in the fiber ribbon by summing the two-photon counts, and dividing by the sum of the singles).

\begin{figure}
\includegraphics[scale=.4]{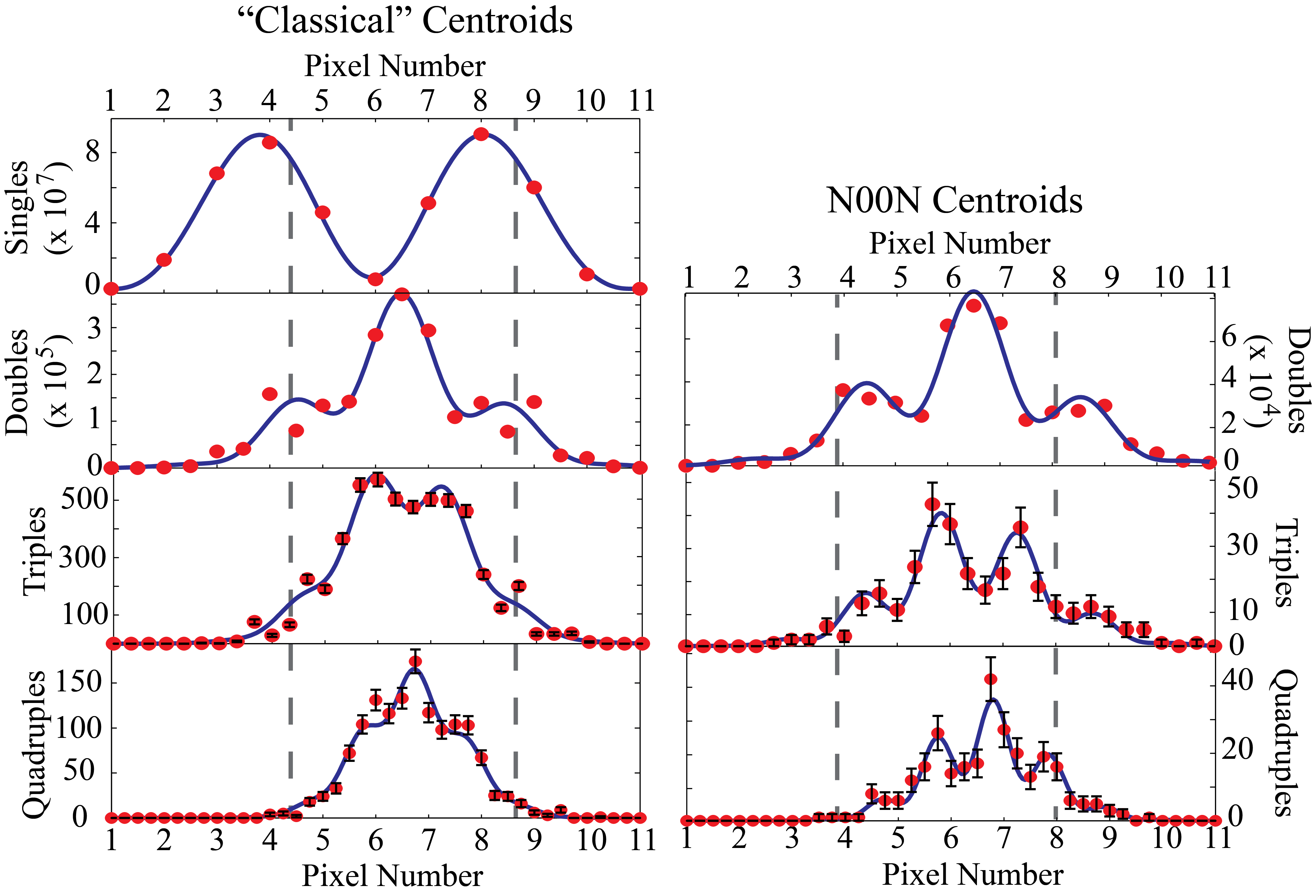}
\caption{{\bf Experimentally measured centroids for one to four photons:}  The first column is the result of a centroid measurement performed on classical light, and the second column the result for N00N states.  The circles are the measured data, and the solid curves are fits from which the visibility is extracted.  The total number of counts is plotted.  The two-, and three-photon N00N-state data were acquired for 66 minutes, and the four-photon N00N-state data were acquired for 57 hours.  To take the classical data, the down-converted light was blocked and the laser intensity was turned up.  Then the one-, two-, and three-photon classical centroid measurements were made in 10 minutes, and the four-photon measurements in 45 minutes.  The error bars are calculated assuming Poisonian counting statistics, and are not shown for the one- and two-photon data as they are much smaller than the circles.  The vertical dashed lines indicate the period of the classical interference pattern.}
\end{figure}

We run the N00N-state source in the two configurations (described above) to take data.  In configuration 1, we observed 150 two-photon N00N states per second and 328 three-photon N00N states in 66 minutes, using the fiber ribbon. In configuration 2, we measured 284 four-photon N00N states in 57 hours.  The resulting OCMs are plotted in the second column of figure 3 (circles).  The three panels, from top to bottom, are the two-, three-, and four-photon OCMs.  A sinusoid with a Gaussian envelope is fitted to the data (solid lines), and the visibility extracted from the sinusoidal part of the fit.  In the fits to the two- to four-photon data, the period of the N-photon centroid is constrained to be the single-photon period divided by N. The resulting visibilities are $49\pm4\%$, $44\pm5\%$ and $41\pm6\%$, for the two-, three-, and four-photon OCMs respectively.  The three- and four- photon visibilities are well in excess of the classical limits of $25\%$ and $12.5\%$, surpassing them by 3.6 and 4.7 standard deviations, respectively.  To account for imperfections in the source (arising from imperfect coupling of the down-converted light), accidental counts were subtracted from the OCM data (explained in the Supplemental Material).  The OCM visibilities, after subtracting these accidental counts, are $65\pm4\%$, $61\pm6\%$ and $59\pm8\%$, for the two-, three-, and four-photon OCMs, respectively.  The accidental-corrected visibilities are better than the classical limits by 3.8, 6.0 and 5.8 standard deviations, respectively.  For comparison, we experimentally perform an OCM on classical light (done by blocking the DC light).  These results are plotted in the first column of figure 3.  These visibilities, again extracted by fitting to the data, are $44\pm9\%$, $18\pm4\%$ and $14\pm4\%$, agreeing well with the expected exponential decay. The increase in the number of ``effective pixels'' is also evident in these data; the singles signal has 11 points, while the four-photon data has 44.  In practical terms, this means an array with fixed spacing could be used to perform an OCM with a N00N state of arbitrary N, unlike the standard N-photon-absorbing scheme which would require the pixel spacing to scale as $\frac{1}{N}$.

\begin{figure}
\includegraphics[scale=.5]{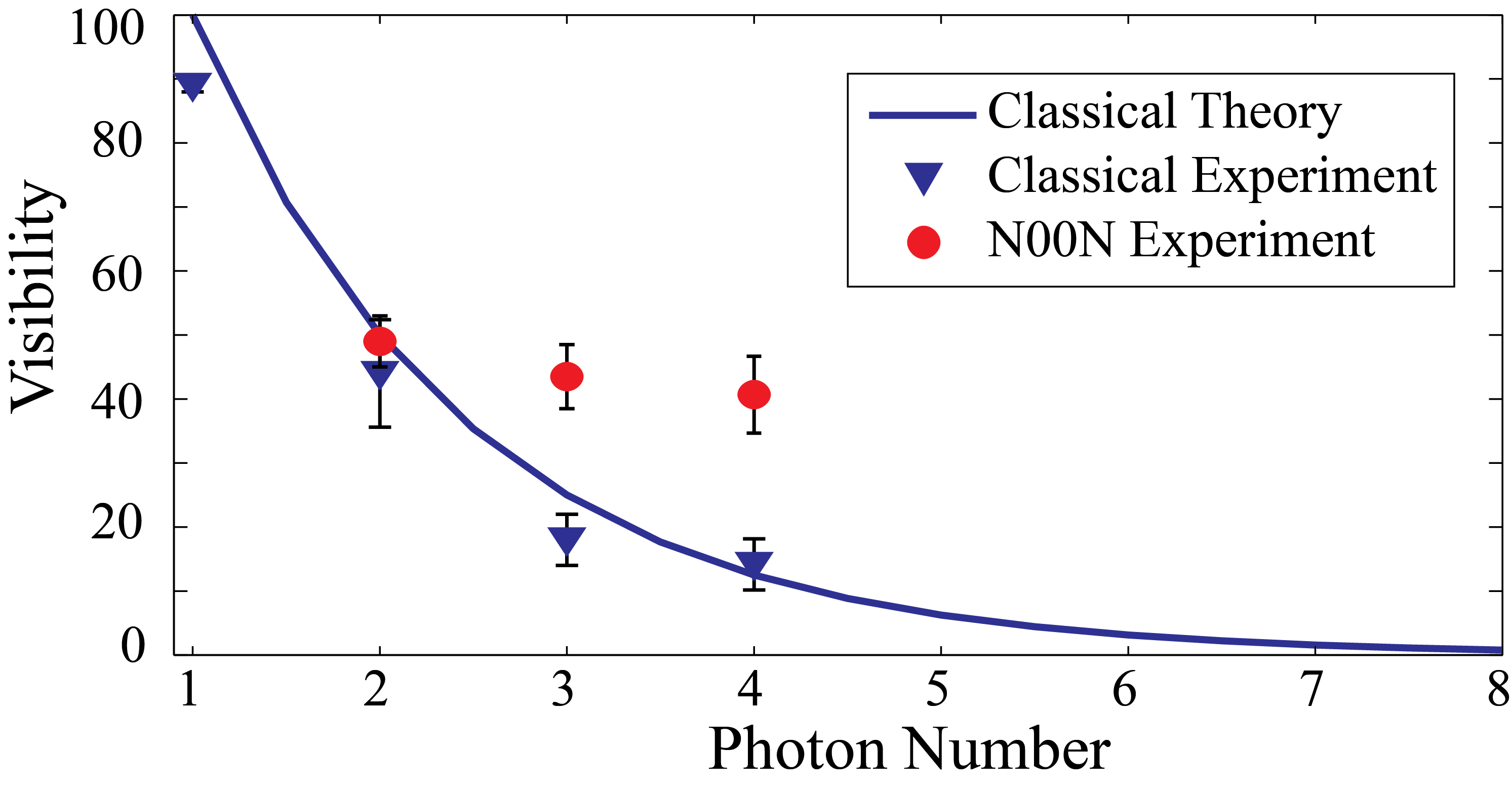}
\caption{{\bf Scaling of visibility with photon number:} A plot of the visibility of the centroid pattern versus photon number. The blue curve is the visibility which is predicted for the classical centroid measurement, the blue triangles are the experientially measured classical visibilities, and the red circles are the measured visibilities of the quantum optical centroid measurement.}
\end{figure}

To study the scalability of our experiment, we plot the various visibilities versus photon number in figure 4.  The solid curve is the theoretical prediction for the visibility of the classical OCM signal.  The experimentally measured classical visibilities (triangles) agree well with this curve, demonstrating a clear exponential decay.  The visibility of the OCM signal performed on a N00N state should be constant (in principle it should be $1.0$).  Although the measured raw visibilities (circles) are not perfect, they are nearly independent of N (this trend is also evident in the accidental-corrected visibilities, plotted in the Supplemental Material).  These measurements demonstrate the scalability of our experiment -- the N00N visibilities remain constant while the classical visibilities decrease exponentially.  This indicates that, given a brighter N00N source, our experimental techniques could be used to measure high-N00N super-resolution signals with high visibility.

In conclusion, we have measured spatial interference patterns of two-, three-, and four-photon N00N states using an 11-detector multiplexed measurement (the OCM).  Our N00N-state source can produce high-fidelity, high-N N00N states \cite{hofmann_high-photon-number_2007, afek_high-noon_2010, rosen_sub-rayleigh_2012}, and we have shown that the N00N state OCM signal displays N-fold super-resolution with a visibility that is nearly independent of N. Thus, our implementation is naturally applicable to higher photon numbers.  This could be achieved using several exciting new technologies which are continually advancing, such as high-efficiency single-photon detectors \cite{rosenberg_noise-free_2005, lita_counting_2008}, high-fill-factor single-photon detector arrays \cite{ghioni_progress_2007}, and brighter down-conversion sources with high coupling efficiency\cite{ramelow_highly_2013, giustina_bell_2013, christensen_detection-loophole-free_2013}.  Our experiment demonstrates a practical implementation which overcomes the problem of efficiently detecting entangled N-photon states -- a fundamental challenge to practical quantum metrology -- and could open the door for Heisenberg-limited phase detection and real quantum imaging.

We thank Amr Helmy and Peter Herman discussions during the design of our experiment, in particular for suggesting the use of a fiber ribbon.  We are very thankful for Alan Stummer's help, designing and building our 11-channel coincidence circuit.  We acknowledge support from the Natural Sciences and Engineering Research Council of Canada and the Canadian Institute for Advanced Research. RO acknowledges additional support from the Yamada Science Foundation.

\bibliography{centroid_DRAFT9.bbl}

\begin{thebibliography}{25}%
\makeatletter
\providecommand \@ifxundefined [1]{%
 \@ifx{#1\undefined}
}%
\providecommand \@ifnum [1]{%
 \ifnum #1\expandafter \@firstoftwo
 \else \expandafter \@secondoftwo
 \fi
}%
\providecommand \@ifx [1]{%
 \ifx #1\expandafter \@firstoftwo
 \else \expandafter \@secondoftwo
 \fi
}%
\providecommand \natexlab [1]{#1}%
\providecommand \enquote  [1]{``#1''}%
\providecommand \bibnamefont  [1]{#1}%
\providecommand \bibfnamefont [1]{#1}%
\providecommand \citenamefont [1]{#1}%
\providecommand \href@noop [0]{\@secondoftwo}%
\providecommand \href [0]{\begingroup \@sanitize@url \@href}%
\providecommand \@href[1]{\@@startlink{#1}\@@href}%
\providecommand \@@href[1]{\endgroup#1\@@endlink}%
\providecommand \@sanitize@url [0]{\catcode `\\12\catcode `\$12\catcode
  `\&12\catcode `\#12\catcode `\^12\catcode `\_12\catcode `\%12\relax}%
\providecommand \@@startlink[1]{}%
\providecommand \@@endlink[0]{}%
\providecommand \url  [0]{\begingroup\@sanitize@url \@url }%
\providecommand \@url [1]{\endgroup\@href {#1}{\urlprefix }}%
\providecommand \urlprefix  [0]{URL }%
\providecommand \Eprint [0]{\href }%
\providecommand \doibase [0]{http://dx.doi.org/}%
\providecommand \selectlanguage [0]{\@gobble}%
\providecommand \bibinfo  [0]{\@secondoftwo}%
\providecommand \bibfield  [0]{\@secondoftwo}%
\providecommand \translation [1]{[#1]}%
\providecommand \BibitemOpen [0]{}%
\providecommand \bibitemStop [0]{}%
\providecommand \bibitemNoStop [0]{.\EOS\space}%
\providecommand \EOS [0]{\spacefactor3000\relax}%
\providecommand \BibitemShut  [1]{\csname bibitem#1\endcsname}%
\let\auto@bib@innerbib\@empty
%</preamble>
\bibitem [{\citenamefont {Rayleigh}(1879)}]{lord_rayleigh_investigations_1879}%
  \BibitemOpen
  \bibfield  {author} {\bibinfo {author} {\bibfnamefont {L.}~\bibnamefont
  {Rayleigh}},\ }\href {\doibase 10.1080/14786447908639684} {\bibfield
  {journal} {\bibinfo  {journal} {Philosophical Magazine Series 5}\ }\textbf
  {\bibinfo {volume} {8}},\ \bibinfo {pages} {261} (\bibinfo {year}
  {1879})}\BibitemShut {NoStop}%
\bibitem [{\citenamefont {Boto}\ \emph {et~al.}(2000)\citenamefont {Boto},
  \citenamefont {Kok}, \citenamefont {Abrams}, \citenamefont {Braunstein},
  \citenamefont {Williams},\ and\ \citenamefont {Dowling}}]{boto_quantum_2000}%
  \BibitemOpen
  \bibfield  {author} {\bibinfo {author} {\bibfnamefont {A.~N.}\ \bibnamefont
  {Boto}}, \bibinfo {author} {\bibfnamefont {P.}~\bibnamefont {Kok}}, \bibinfo
  {author} {\bibfnamefont {D.~S.}\ \bibnamefont {Abrams}}, \bibinfo {author}
  {\bibfnamefont {S.~L.}\ \bibnamefont {Braunstein}}, \bibinfo {author}
  {\bibfnamefont {C.~P.}\ \bibnamefont {Williams}}, \ and\ \bibinfo {author}
  {\bibfnamefont {J.~P.}\ \bibnamefont {Dowling}},\ }\href {\doibase
  10.1103/PhysRevLett.85.2733} {\bibfield  {journal} {\bibinfo  {journal}
  {Physical Review Letters}\ }\textbf {\bibinfo {volume} {85}},\ \bibinfo
  {pages} {2733} (\bibinfo {year} {2000})}\BibitemShut {NoStop}%
\bibitem [{\citenamefont {Kok}\ \emph {et~al.}(2001)\citenamefont {Kok},
  \citenamefont {Boto}, \citenamefont {Abrams}, \citenamefont {Williams},
  \citenamefont {Braunstein},\ and\ \citenamefont
  {Dowling}}]{kok_quantum-interferometric_2001}%
  \BibitemOpen
  \bibfield  {author} {\bibinfo {author} {\bibfnamefont {P.}~\bibnamefont
  {Kok}}, \bibinfo {author} {\bibfnamefont {A.~N.}\ \bibnamefont {Boto}},
  \bibinfo {author} {\bibfnamefont {D.~S.}\ \bibnamefont {Abrams}}, \bibinfo
  {author} {\bibfnamefont {C.~P.}\ \bibnamefont {Williams}}, \bibinfo {author}
  {\bibfnamefont {S.~L.}\ \bibnamefont {Braunstein}}, \ and\ \bibinfo {author}
  {\bibfnamefont {J.~P.}\ \bibnamefont {Dowling}},\ }\href {\doibase
  10.1103/PhysRevA.63.063407} {\bibfield  {journal} {\bibinfo  {journal}
  {Physical Review A}\ }\textbf {\bibinfo {volume} {63}},\ \bibinfo {pages}
  {063407} (\bibinfo {year} {2001})}\BibitemShut {NoStop}%
\bibitem [{\citenamefont {Mitchell}\ \emph {et~al.}(2004)\citenamefont
  {Mitchell}, \citenamefont {Lundeen},\ and\ \citenamefont
  {Steinberg}}]{mitchell_super-resolving_2004}%
  \BibitemOpen
  \bibfield  {author} {\bibinfo {author} {\bibfnamefont {M.~W.}\ \bibnamefont
  {Mitchell}}, \bibinfo {author} {\bibfnamefont {J.~S.}\ \bibnamefont
  {Lundeen}}, \ and\ \bibinfo {author} {\bibfnamefont {A.~M.}\ \bibnamefont
  {Steinberg}},\ }\href {\doibase 10.1038/nature02493} {\bibfield  {journal}
  {\bibinfo  {journal} {Nature}\ }\textbf {\bibinfo {volume} {429}},\ \bibinfo
  {pages} {161} (\bibinfo {year} {2004})}\BibitemShut {NoStop}%
\bibitem [{\citenamefont {Walther}\ \emph {et~al.}(2004)\citenamefont
  {Walther}, \citenamefont {Pan}, \citenamefont {Aspelmeyer}, \citenamefont
  {Ursin}, \citenamefont {Gasparoni},\ and\ \citenamefont
  {Zeilinger}}]{walther_broglie_2004}%
  \BibitemOpen
  \bibfield  {author} {\bibinfo {author} {\bibfnamefont {P.}~\bibnamefont
  {Walther}}, \bibinfo {author} {\bibfnamefont {J.-W.}\ \bibnamefont {Pan}},
  \bibinfo {author} {\bibfnamefont {M.}~\bibnamefont {Aspelmeyer}}, \bibinfo
  {author} {\bibfnamefont {R.}~\bibnamefont {Ursin}}, \bibinfo {author}
  {\bibfnamefont {S.}~\bibnamefont {Gasparoni}}, \ and\ \bibinfo {author}
  {\bibfnamefont {A.}~\bibnamefont {Zeilinger}},\ }\href {\doibase
  10.1038/nature02552} {\bibfield  {journal} {\bibinfo  {journal} {Nature}\
  }\textbf {\bibinfo {volume} {429}},\ \bibinfo {pages} {158} (\bibinfo {year}
  {2004})}\BibitemShut {NoStop}%
\bibitem [{\citenamefont {Steuernagel}(2004)}]{steuernagel_concentration_2004}%
  \BibitemOpen
  \bibfield  {author} {\bibinfo {author} {\bibfnamefont {O.}~\bibnamefont
  {Steuernagel}},\ }\href {\doibase 10.1088/1464-4266/6/6/021} {\bibfield
  {journal} {\bibinfo  {journal} {Journal of Optics B: Quantum and
  Semiclassical Optics}\ }\textbf {\bibinfo {volume} {6}},\ \bibinfo {pages}
  {S606} (\bibinfo {year} {2004})}\BibitemShut {NoStop}%
\bibitem [{\citenamefont {Tsang}(2007)}]{tsang_relationship_2007}%
  \BibitemOpen
  \bibfield  {author} {\bibinfo {author} {\bibfnamefont {M.}~\bibnamefont
  {Tsang}},\ }\href {\doibase 10.1103/PhysRevA.75.043813} {\bibfield  {journal}
  {\bibinfo  {journal} {Physical Review A}\ }\textbf {\bibinfo {volume} {75}},\
  \bibinfo {pages} {043813} (\bibinfo {year} {2007})}\BibitemShut {NoStop}%
\bibitem [{\citenamefont {Tsang}(2009)}]{tsang_quantum_2009}%
  \BibitemOpen
  \bibfield  {author} {\bibinfo {author} {\bibfnamefont {M.}~\bibnamefont
  {Tsang}},\ }\href {\doibase 10.1103/PhysRevLett.102.253601} {\bibfield
  {journal} {\bibinfo  {journal} {Physical Review Letters}\ }\textbf {\bibinfo
  {volume} {102}},\ \bibinfo {pages} {253601} (\bibinfo {year}
  {2009})}\BibitemShut {NoStop}%
\bibitem [{\citenamefont {Shin}\ \emph {et~al.}(2011)\citenamefont {Shin},
  \citenamefont {Chan}, \citenamefont {Chang},\ and\ \citenamefont
  {Boyd}}]{shin_quantum_2011}%
  \BibitemOpen
  \bibfield  {author} {\bibinfo {author} {\bibfnamefont {H.}~\bibnamefont
  {Shin}}, \bibinfo {author} {\bibfnamefont {K.~W.~C.}\ \bibnamefont {Chan}},
  \bibinfo {author} {\bibfnamefont {H.~J.}\ \bibnamefont {Chang}}, \ and\
  \bibinfo {author} {\bibfnamefont {R.~W.}\ \bibnamefont {Boyd}},\ }\href
  {\doibase 10.1103/PhysRevLett.107.083603} {\bibfield  {journal} {\bibinfo
  {journal} {Physical Review Letters}\ }\textbf {\bibinfo {volume} {107}},\
  \bibinfo {pages} {083603} (\bibinfo {year} {2011})}\BibitemShut {NoStop}%
\bibitem [{\citenamefont {{D'Angelo}}\ \emph {et~al.}(2001)\citenamefont
  {{D'Angelo}}, \citenamefont {Chekhova},\ and\ \citenamefont
  {Shih}}]{dangelo_two-photon_2001}%
  \BibitemOpen
  \bibfield  {author} {\bibinfo {author} {\bibfnamefont {M.}~\bibnamefont
  {{D'Angelo}}}, \bibinfo {author} {\bibfnamefont {M.~V.}\ \bibnamefont
  {Chekhova}}, \ and\ \bibinfo {author} {\bibfnamefont {Y.}~\bibnamefont
  {Shih}},\ }\href {\doibase 10.1103/PhysRevLett.87.013602} {\bibfield
  {journal} {\bibinfo  {journal} {Physical Review Letters}\ }\textbf {\bibinfo
  {volume} {87}},\ \bibinfo {pages} {013602} (\bibinfo {year}
  {2001})}\BibitemShut {NoStop}%
\bibitem [{\citenamefont {Kawabe}\ \emph {et~al.}(2007)\citenamefont {Kawabe},
  \citenamefont {Fujiwara}, \citenamefont {Okamoto}, \citenamefont {Sasaki},\
  and\ \citenamefont {Takeuchi}}]{kawabe_quantum_2007}%
  \BibitemOpen
  \bibfield  {author} {\bibinfo {author} {\bibfnamefont {Y.}~\bibnamefont
  {Kawabe}}, \bibinfo {author} {\bibfnamefont {H.}~\bibnamefont {Fujiwara}},
  \bibinfo {author} {\bibfnamefont {R.}~\bibnamefont {Okamoto}}, \bibinfo
  {author} {\bibfnamefont {K.}~\bibnamefont {Sasaki}}, \ and\ \bibinfo {author}
  {\bibfnamefont {S.}~\bibnamefont {Takeuchi}},\ }\href {\doibase
  10.1364/OE.15.014244} {\bibfield  {journal} {\bibinfo  {journal} {Optics
  Express}\ }\textbf {\bibinfo {volume} {15}},\ \bibinfo {pages} {14244}
  (\bibinfo {year} {2007})}\BibitemShut {NoStop}%
\bibitem [{\citenamefont {Kim}\ \emph {et~al.}(2011)\citenamefont {Kim},
  \citenamefont {Kwon}, \citenamefont {Lee}, \citenamefont {Lee}, \citenamefont
  {Kim}, \citenamefont {Choi}, \citenamefont {Park},\ and\ \citenamefont
  {Kim}}]{kim_observation_2011}%
  \BibitemOpen
  \bibfield  {author} {\bibinfo {author} {\bibfnamefont {Y.-S.}\ \bibnamefont
  {Kim}}, \bibinfo {author} {\bibfnamefont {O.}~\bibnamefont {Kwon}}, \bibinfo
  {author} {\bibfnamefont {S.~M.}\ \bibnamefont {Lee}}, \bibinfo {author}
  {\bibfnamefont {J.-C.}\ \bibnamefont {Lee}}, \bibinfo {author} {\bibfnamefont
  {H.}~\bibnamefont {Kim}}, \bibinfo {author} {\bibfnamefont {S.-K.}\
  \bibnamefont {Choi}}, \bibinfo {author} {\bibfnamefont {H.~S.}\ \bibnamefont
  {Park}}, \ and\ \bibinfo {author} {\bibfnamefont {Y.-H.}\ \bibnamefont
  {Kim}},\ }\href {\doibase 10.1364/OE.19.024957} {\bibfield  {journal}
  {\bibinfo  {journal} {Optics Express}\ }\textbf {\bibinfo {volume} {19}},\
  \bibinfo {pages} {24957} (\bibinfo {year} {2011})}\BibitemShut {NoStop}%
\bibitem [{\citenamefont {Nagata}\ \emph {et~al.}(2007)\citenamefont {Nagata},
  \citenamefont {Okamoto}, \citenamefont {{O'Brien}}, \citenamefont {Sasaki},\
  and\ \citenamefont {Takeuchi}}]{nagata_beating_2007}%
  \BibitemOpen
  \bibfield  {author} {\bibinfo {author} {\bibfnamefont {T.}~\bibnamefont
  {Nagata}}, \bibinfo {author} {\bibfnamefont {R.}~\bibnamefont {Okamoto}},
  \bibinfo {author} {\bibfnamefont {J.~L.}\ \bibnamefont {{O'Brien}}}, \bibinfo
  {author} {\bibfnamefont {K.}~\bibnamefont {Sasaki}}, \ and\ \bibinfo {author}
  {\bibfnamefont {S.}~\bibnamefont {Takeuchi}},\ }\href {\doibase
  10.1126/science.1138007} {\bibfield  {journal} {\bibinfo  {journal}
  {Science}\ }\textbf {\bibinfo {volume} {316}},\ \bibinfo {pages} {726}
  (\bibinfo {year} {2007})}\BibitemShut {NoStop}%
\bibitem [{\citenamefont {Shalm}\ \emph {et~al.}(2009)\citenamefont {Shalm},
  \citenamefont {Adamson},\ and\ \citenamefont
  {Steinberg}}]{shalm_squeezing_2009}%
  \BibitemOpen
  \bibfield  {author} {\bibinfo {author} {\bibfnamefont {L.~K.}\ \bibnamefont
  {Shalm}}, \bibinfo {author} {\bibfnamefont {R.~B.~A.}\ \bibnamefont
  {Adamson}}, \ and\ \bibinfo {author} {\bibfnamefont {A.~M.}\ \bibnamefont
  {Steinberg}},\ }\href {\doibase 10.1038/nature07624} {\bibfield  {journal}
  {\bibinfo  {journal} {Nature}\ }\textbf {\bibinfo {volume} {457}},\ \bibinfo
  {pages} {67} (\bibinfo {year} {2009})}\BibitemShut {NoStop}%
\bibitem [{\citenamefont {Hofmann}\ and\ \citenamefont
  {Ono}(2007)}]{hofmann_high-photon-number_2007}%
  \BibitemOpen
  \bibfield  {author} {\bibinfo {author} {\bibfnamefont {H.~F.}\ \bibnamefont
  {Hofmann}}\ and\ \bibinfo {author} {\bibfnamefont {T.}~\bibnamefont {Ono}},\
  }\href {\doibase 10.1103/PhysRevA.76.031806} {\bibfield  {journal} {\bibinfo
  {journal} {Physical Review A}\ }\textbf {\bibinfo {volume} {76}},\ \bibinfo
  {pages} {031806} (\bibinfo {year} {2007})}\BibitemShut {NoStop}%
\bibitem [{\citenamefont {Afek}\ \emph {et~al.}(2010)\citenamefont {Afek},
  \citenamefont {Ambar},\ and\ \citenamefont
  {Silberberg}}]{afek_high-noon_2010}%
  \BibitemOpen
  \bibfield  {author} {\bibinfo {author} {\bibfnamefont {I.}~\bibnamefont
  {Afek}}, \bibinfo {author} {\bibfnamefont {O.}~\bibnamefont {Ambar}}, \ and\
  \bibinfo {author} {\bibfnamefont {Y.}~\bibnamefont {Silberberg}},\ }\href
  {\doibase 10.1126/science.1188172} {\bibfield  {journal} {\bibinfo  {journal}
  {Science}\ }\textbf {\bibinfo {volume} {328}},\ \bibinfo {pages} {879}
  (\bibinfo {year} {2010})}\BibitemShut {NoStop}%
\bibitem [{\citenamefont {Nasr}\ \emph {et~al.}(2003)\citenamefont {Nasr},
  \citenamefont {Saleh}, \citenamefont {Sergienko},\ and\ \citenamefont
  {Teich}}]{nasr_demonstration_2003}%
  \BibitemOpen
  \bibfield  {author} {\bibinfo {author} {\bibfnamefont {M.~B.}\ \bibnamefont
  {Nasr}}, \bibinfo {author} {\bibfnamefont {B.~E.~A.}\ \bibnamefont {Saleh}},
  \bibinfo {author} {\bibfnamefont {A.~V.}\ \bibnamefont {Sergienko}}, \ and\
  \bibinfo {author} {\bibfnamefont {M.~C.}\ \bibnamefont {Teich}},\ }\href
  {\doibase 10.1103/PhysRevLett.91.083601} {\bibfield  {journal} {\bibinfo
  {journal} {Physical Review Letters}\ }\textbf {\bibinfo {volume} {91}},\
  \bibinfo {pages} {083601} (\bibinfo {year} {2003})}\BibitemShut {NoStop}%
\bibitem [{\citenamefont {Gulfam}\ and\ \citenamefont
  {Evers}(2013)}]{gulfam_numerical_2013}%
  \BibitemOpen
  \bibfield  {author} {\bibinfo {author} {\bibfnamefont {Q.}~\bibnamefont
  {Gulfam}}\ and\ \bibinfo {author} {\bibfnamefont {J.}~\bibnamefont {Evers}},\
  }\href {\doibase 10.1103/PhysRevA.87.023804} {\bibfield  {journal} {\bibinfo
  {journal} {Physical Review A}\ }\textbf {\bibinfo {volume} {87}},\ \bibinfo
  {pages} {023804} (\bibinfo {year} {2013})}\BibitemShut {NoStop}%
\bibitem [{\citenamefont {Rosen}\ \emph {et~al.}(2012)\citenamefont {Rosen},
  \citenamefont {Afek}, \citenamefont {Israel}, \citenamefont {Ambar},\ and\
  \citenamefont {Silberberg}}]{rosen_sub-rayleigh_2012}%
  \BibitemOpen
  \bibfield  {author} {\bibinfo {author} {\bibfnamefont {S.}~\bibnamefont
  {Rosen}}, \bibinfo {author} {\bibfnamefont {I.}~\bibnamefont {Afek}},
  \bibinfo {author} {\bibfnamefont {Y.}~\bibnamefont {Israel}}, \bibinfo
  {author} {\bibfnamefont {O.}~\bibnamefont {Ambar}}, \ and\ \bibinfo {author}
  {\bibfnamefont {Y.}~\bibnamefont {Silberberg}},\ }\href {\doibase
  10.1103/PhysRevLett.109.103602} {\bibfield  {journal} {\bibinfo  {journal}
  {Physical Review Letters}\ }\textbf {\bibinfo {volume} {109}},\ \bibinfo
  {pages} {103602} (\bibinfo {year} {2012})}\BibitemShut {NoStop}%
\bibitem [{\citenamefont {Rosenberg}\ \emph {et~al.}(2005)\citenamefont
  {Rosenberg}, \citenamefont {Lita}, \citenamefont {Miller},\ and\
  \citenamefont {Nam}}]{rosenberg_noise-free_2005}%
  \BibitemOpen
  \bibfield  {author} {\bibinfo {author} {\bibfnamefont {D.}~\bibnamefont
  {Rosenberg}}, \bibinfo {author} {\bibfnamefont {A.~E.}\ \bibnamefont {Lita}},
  \bibinfo {author} {\bibfnamefont {A.~J.}\ \bibnamefont {Miller}}, \ and\
  \bibinfo {author} {\bibfnamefont {S.~W.}\ \bibnamefont {Nam}},\ }\href
  {\doibase 10.1103/PhysRevA.71.061803} {\bibfield  {journal} {\bibinfo
  {journal} {Physical Review A}\ }\textbf {\bibinfo {volume} {71}},\ \bibinfo
  {pages} {061803} (\bibinfo {year} {2005})}\BibitemShut {NoStop}%
\bibitem [{\citenamefont {Lita}\ \emph {et~al.}(2008)\citenamefont {Lita},
  \citenamefont {Miller},\ and\ \citenamefont {Nam}}]{lita_counting_2008}%
  \BibitemOpen
  \bibfield  {author} {\bibinfo {author} {\bibfnamefont {A.~E.}\ \bibnamefont
  {Lita}}, \bibinfo {author} {\bibfnamefont {A.~J.}\ \bibnamefont {Miller}}, \
  and\ \bibinfo {author} {\bibfnamefont {S.~W.}\ \bibnamefont {Nam}},\ }\href
  {\doibase 10.1364/OE.16.003032} {\bibfield  {journal} {\bibinfo  {journal}
  {Optics Express}\ }\textbf {\bibinfo {volume} {16}},\ \bibinfo {pages} {3032}
  (\bibinfo {year} {2008})}\BibitemShut {NoStop}%
\bibitem [{\citenamefont {Ghioni}\ \emph {et~al.}(2007)\citenamefont {Ghioni},
  \citenamefont {Gulinatti}, \citenamefont {Rech}, \citenamefont {Zappa},\ and\
  \citenamefont {Cova}}]{ghioni_progress_2007}%
  \BibitemOpen
  \bibfield  {author} {\bibinfo {author} {\bibfnamefont {M.}~\bibnamefont
  {Ghioni}}, \bibinfo {author} {\bibfnamefont {A.}~\bibnamefont {Gulinatti}},
  \bibinfo {author} {\bibfnamefont {I.}~\bibnamefont {Rech}}, \bibinfo {author}
  {\bibfnamefont {F.}~\bibnamefont {Zappa}}, \ and\ \bibinfo {author}
  {\bibfnamefont {S.}~\bibnamefont {Cova}},\ }\href {\doibase
  10.1109/JSTQE.2007.902088} {\bibfield  {journal} {\bibinfo  {journal} {{IEEE}
  Journal of Selected Topics in Quantum Electronics}\ }\textbf {\bibinfo
  {volume} {13}},\ \bibinfo {pages} {852} (\bibinfo {year} {2007})}\BibitemShut
  {NoStop}%
\bibitem [{\citenamefont {Ramelow}\ \emph {et~al.}(2013)\citenamefont
  {Ramelow}, \citenamefont {Mech}, \citenamefont {Giustina}, \citenamefont
  {Groblacher}, \citenamefont {Wieczorek}, \citenamefont {Beyer}, \citenamefont
  {Lita}, \citenamefont {Calkins}, \citenamefont {Gerrits}, \citenamefont
  {Nam}, \citenamefont {Zeilinger},\ and\ \citenamefont
  {Ursin}}]{ramelow_highly_2013}%
  \BibitemOpen
  \bibfield  {author} {\bibinfo {author} {\bibfnamefont {S.}~\bibnamefont
  {Ramelow}}, \bibinfo {author} {\bibfnamefont {A.}~\bibnamefont {Mech}},
  \bibinfo {author} {\bibfnamefont {M.}~\bibnamefont {Giustina}}, \bibinfo
  {author} {\bibfnamefont {S.}~\bibnamefont {Groblacher}}, \bibinfo {author}
  {\bibfnamefont {W.}~\bibnamefont {Wieczorek}}, \bibinfo {author}
  {\bibfnamefont {J.}~\bibnamefont {Beyer}}, \bibinfo {author} {\bibfnamefont
  {A.}~\bibnamefont {Lita}}, \bibinfo {author} {\bibfnamefont {B.}~\bibnamefont
  {Calkins}}, \bibinfo {author} {\bibfnamefont {T.}~\bibnamefont {Gerrits}},
  \bibinfo {author} {\bibfnamefont {S.~W.}\ \bibnamefont {Nam}}, \bibinfo
  {author} {\bibfnamefont {A.}~\bibnamefont {Zeilinger}}, \ and\ \bibinfo
  {author} {\bibfnamefont {R.}~\bibnamefont {Ursin}},\ }\href {\doibase
  10.1364/OE.21.006707} {\bibfield  {journal} {\bibinfo  {journal} {Optics
  Express}\ }\textbf {\bibinfo {volume} {21}},\ \bibinfo {pages} {6707}
  (\bibinfo {year} {2013})}\BibitemShut {NoStop}%
\bibitem [{\citenamefont {Giustina}\ \emph {et~al.}(2013)\citenamefont
  {Giustina}, \citenamefont {Mech}, \citenamefont {Ramelow}, \citenamefont
  {Wittmann}, \citenamefont {Kofler}, \citenamefont {Beyer}, \citenamefont
  {Lita}, \citenamefont {Calkins}, \citenamefont {Gerrits}, \citenamefont
  {Nam}, \citenamefont {Ursin},\ and\ \citenamefont
  {Zeilinger}}]{giustina_bell_2013}%
  \BibitemOpen
  \bibfield  {author} {\bibinfo {author} {\bibfnamefont {M.}~\bibnamefont
  {Giustina}}, \bibinfo {author} {\bibfnamefont {A.}~\bibnamefont {Mech}},
  \bibinfo {author} {\bibfnamefont {S.}~\bibnamefont {Ramelow}}, \bibinfo
  {author} {\bibfnamefont {B.}~\bibnamefont {Wittmann}}, \bibinfo {author}
  {\bibfnamefont {J.}~\bibnamefont {Kofler}}, \bibinfo {author} {\bibfnamefont
  {J.}~\bibnamefont {Beyer}}, \bibinfo {author} {\bibfnamefont
  {A.}~\bibnamefont {Lita}}, \bibinfo {author} {\bibfnamefont {B.}~\bibnamefont
  {Calkins}}, \bibinfo {author} {\bibfnamefont {T.}~\bibnamefont {Gerrits}},
  \bibinfo {author} {\bibfnamefont {S.~W.}\ \bibnamefont {Nam}}, \bibinfo
  {author} {\bibfnamefont {R.}~\bibnamefont {Ursin}}, \ and\ \bibinfo {author}
  {\bibfnamefont {A.}~\bibnamefont {Zeilinger}},\ }\href {\doibase
  10.1038/nature12012} {\bibfield  {journal} {\bibinfo  {journal} {Nature}\
  }\textbf {\bibinfo {volume} {497}},\ \bibinfo {pages} {227} (\bibinfo {year}
  {2013})}\BibitemShut {NoStop}%
\bibitem [{\citenamefont {Christensen}\ \emph {et~al.}(2013)\citenamefont
  {Christensen}, \citenamefont {{McCusker}}, \citenamefont {Altepeter},
  \citenamefont {Calkins}, \citenamefont {Gerrits}, \citenamefont {Lita},
  \citenamefont {Miller}, \citenamefont {Shalm}, \citenamefont {Zhang},
  \citenamefont {Nam}, \citenamefont {Brunner}, \citenamefont {Lim},
  \citenamefont {Gisin},\ and\ \citenamefont
  {Kwiat}}]{christensen_detection-loophole-free_2013}%
  \BibitemOpen
  \bibfield  {author} {\bibinfo {author} {\bibfnamefont {B.~G.}\ \bibnamefont
  {Christensen}}, \bibinfo {author} {\bibfnamefont {K.~T.}\ \bibnamefont
  {{McCusker}}}, \bibinfo {author} {\bibfnamefont {J.~B.}\ \bibnamefont
  {Altepeter}}, \bibinfo {author} {\bibfnamefont {B.}~\bibnamefont {Calkins}},
  \bibinfo {author} {\bibfnamefont {T.}~\bibnamefont {Gerrits}}, \bibinfo
  {author} {\bibfnamefont {A.~E.}\ \bibnamefont {Lita}}, \bibinfo {author}
  {\bibfnamefont {A.}~\bibnamefont {Miller}}, \bibinfo {author} {\bibfnamefont
  {L.~K.}\ \bibnamefont {Shalm}}, \bibinfo {author} {\bibfnamefont
  {Y.}~\bibnamefont {Zhang}}, \bibinfo {author} {\bibfnamefont {S.~W.}\
  \bibnamefont {Nam}}, \bibinfo {author} {\bibfnamefont {N.}~\bibnamefont
  {Brunner}}, \bibinfo {author} {\bibfnamefont {C.~C.~W.}\ \bibnamefont {Lim}},
  \bibinfo {author} {\bibfnamefont {N.}~\bibnamefont {Gisin}}, \ and\ \bibinfo
  {author} {\bibfnamefont {P.~G.}\ \bibnamefont {Kwiat}},\ }\href {\doibase
  10.1103/PhysRevLett.111.130406} {\bibfield  {journal} {\bibinfo  {journal}
  {Physical Review Letters}\ }\textbf {\bibinfo {volume} {111}},\ \bibinfo
  {pages} {130406} (\bibinfo {year} {2013})}\BibitemShut {NoStop}%
\end{thebibliography}%
\newpage
\section{Supplemental Information}
\subsection{Visibility of Classical Centroid}
When performing the optical centroid measurement (OCM), the resulting signal, $C(X)$ (where $X$ the centroid coordinate), can be written in terms of the N-photon probability distribution, $P({x_1,\dots,x_N})$ (the probability of the N-photons arriving at positions $x_1$ to $x_N$), as:
\begin{equation}
C(X)=\int dx_1\dots\int dx_N P({x_1,\dots,x_N}) \delta(\frac{x_1+\dots+x_N}{N}-X).
\end{equation}
If the incident light is classical, $P({x_1,\dots,x_N})$ factorizes into $P(x_1)\times\dots\times P(x_N)$.  In this case, $C(X)$ becomes the N-fold convolution of N identical single-photon probability distributions:
\begin{equation}
C_{cl}(X)=(P(x_1)*\dots*P(x_N))(N X).
\end{equation}
The single-photon probability distribution for classical light (of wavelength $\lambda$) interfering at an angle of $\theta$ is $P(x)\propto1+\cos(fx)$ (where $f=(4\pi\sin\frac{\theta}{2})/\lambda$). The Fourier transform of this single-photon probability distribution is $\tilde{P}(\omega)\propto\frac{1}{2}\delta(\omega+f)+\delta(\omega)+\frac{1}{2}\delta(\omega-f)$. Due to the Convolution Theorem (the Fourier transform of the convolution is the product of the Fourier transforms of the individual functions) the Fourier transform of the classical centroid signal, $C_{cl}(X)$, is $\tilde{C}_{cl}(\omega)\propto\frac{1}{2^N}\delta(\omega+f)+\delta(\omega)+\frac{1}{2^N}\delta(\omega-f)$.  So the $C_{cl}(X)$ becomes
\begin{equation}
C_{cl}(X)\propto1+\frac{1}{2^{N-1}}\cos(Nf X).
\end{equation}
The visibility of $C_{cl}(X)$ is $\frac{1}{2^{N-1}}$. Thus, although the centroid signal will exhibit fringes with a frequency times $N$ larger than the classical signal, the visibility of those fringes decreases exponentially.

\subsection{Subtraction of Accidentals}

In our N00N-state source, we combine classical light with down-converted light, and then post-select on detecting N photons. Assuming perfect detectors, when N photons are detected they will be in a state which has high fidelity with the ideal N00N state.  In reality, our detectors have approximately $60\%$ efficiency and photons may be lost before they even impinge on the detectors.  In these conditions, unwanted ``accidental'' counts can occur.  To understand this, consider making a three-photon N00N state.  In this case, the N00N state should arise due to interference between the possibility of detecting three laser photons and no down-converted photons and that of detecting one laser photon and two down-converted photons.  Any other detection event will lower the fidelity with the N00N state.  Since the detectors and the coupling efficiency are imperfect, it is possible to lose one down-converted photon and subsequently detect one down-converted photon and two laser photons.  This accidental event cannot be distinguished from the desired events.

\begin{figure}
\includegraphics[scale=.5]{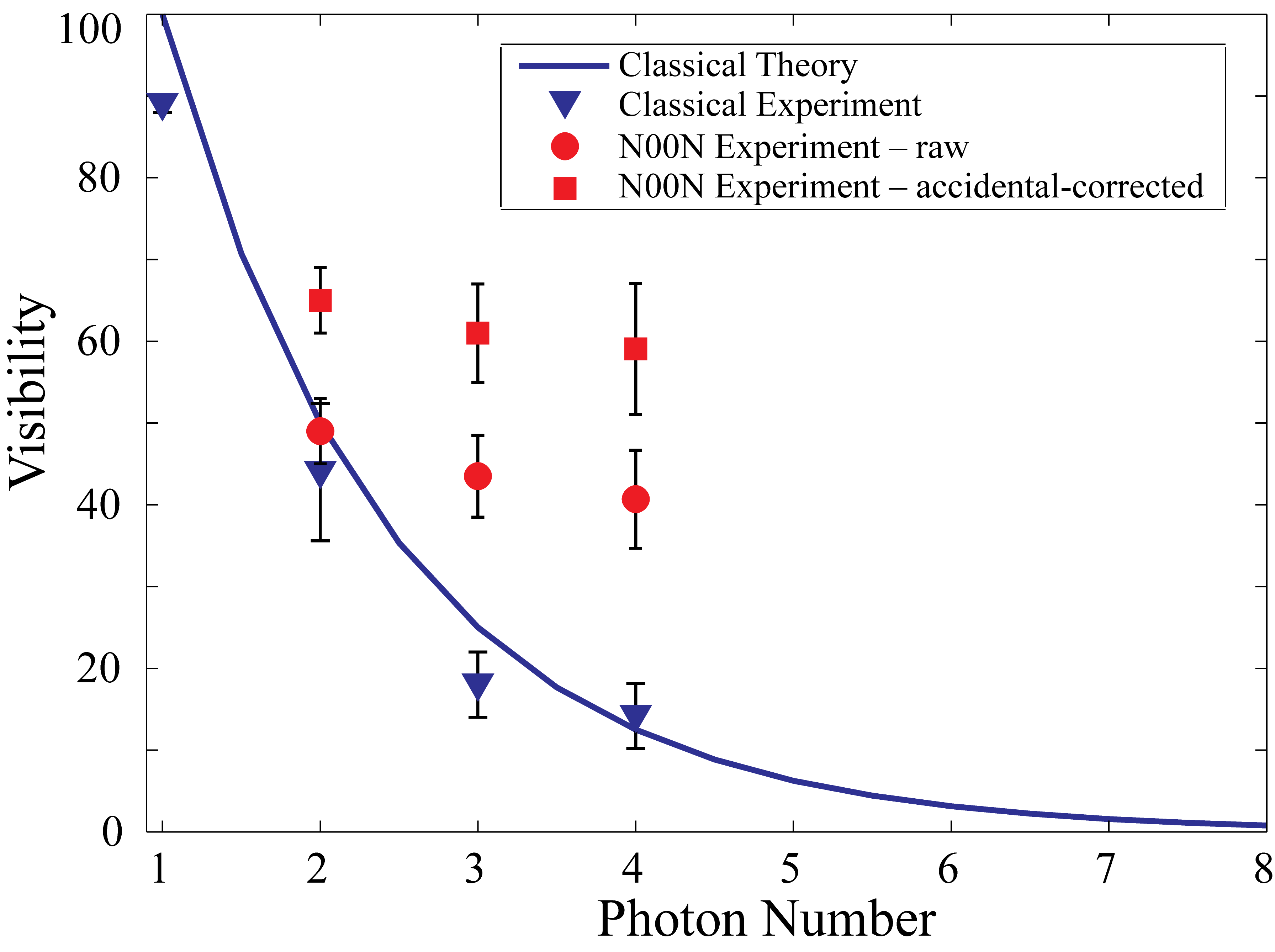}
\caption{{\bf Scaling of visibility with photon number:} A plot of the visibility of the centroid pattern versus photon number. The blue curve is the visibility which is predicted for the classical centroid measurement, and the blue triangles are the experientially measured classical visibilities. The red circles are the measured visibilities of the quantum centroid visibility, and the red squares are the visibilities of the same quantum centroid patterns after being corrected for accidental counts.}
\end{figure}

We estimate the rate of accidental events and subtract them from our counts.  The accidental rate is estimated by blocking the laser and the down-converted beam one at a time, and measuring the singles rate on the 11 pixels in our array for each.  We then calculated the predicted two-, three- and four-photon accidental rates at each centroid position from the singles rates.  In our estimation of the accidental counts, we assumed that the accidental counts were randomly distributed; i.e. if one photon of a down-converted pair is lost its detected twin will not undergo Hong-Ou-Mandel--like interference with the laser photons.  These accidentals are then subtracted from our measured centroid counts.  On average, we find that this increases the visibility roughly from $40\%$ to $60\%$.  The accidental-corrected visibilities are displayed as the red squares in supplementary figure 5, alongside the uncorrected N00N state data (red circles), and the classical visibilities (blue triangles).

\end{document}